\documentclass[%
 aip,
 amsmath,amssymb,
 reprint,%
 floatfix,%
]{revtex4-1}

\usepackage{graphicx}
\usepackage{bm}

\usepackage[utf8]{inputenc}
\usepackage[T1]{fontenc}
\usepackage{mathptmx}
\usepackage{etoolbox}
\usepackage{float}
\usepackage{hyperref}
\providecommand{\collaboration}[1]{\par\noindent\textbf{#1}\par}
\def\@email#1#2{%
 \endgroup
 \patchcmd{\titleblock@produce}
  {\frontmatter@RRAPformat}
  {\frontmatter@RRAPformat{\produce@RRAP{*#1\href{mailto:#2}{#2}}}\frontmatter@RRAPformat}
  {}{}
}%
\makeatother
\begin{document}

\preprint{AIP/123-QED}

\title{Rumor Propagation and Supervision during Confrontation: An 
Importance-Driven SIRQS Network Model}

\author{Wenjie Zhang}
\affiliation{Innovation and Entrepreneurship College (Han Hong College), Southwest University, Chongqing, 400715, China}

\author{Juan Wu}
\affiliation{School of Statistics and Data Science, Xi'an University of Finance and Economics, Xi'an 710100, China}

\author{Minyu Feng}
\affiliation{College of Artificial Intelligence, Southwest University, Chongqing, 400715, China}

\author{Qin Li}
\email{Author to whom correspondence should be addressed: qinli1022@swu.edu.cn}
\affiliation{Business College, Southwest University, Chongqing, 402460, China}

\author{Matja\v{z} Perc }
\email{Author to whom correspondence should be addressed: matjaz.perc@gmail.com}
\affiliation{Faculty of Natural Sciences and Mathematics, University of Maribor,
Koro{\v s}ka cesta 160, 2000 Maribor, Slovenia}
\affiliation{Community Healthcare Center Dr. Adolf Drolc Maribor, Ulica talcev 9,
2000 Maribor, Slovenia}
\affiliation{Department of Physics, Kyung Hee University, 26 Kyungheedae-ro,
Dongdaemun-gu, Seoul 02447, Republic of Korea}
\affiliation{University College, Korea University, 145 Anam-ro, Seongbuk-gu,
Seoul 02841, Republic of Korea}
\collaboration{Corresponding authors: Qin Li and Matja\v{z} Perc.}

\begin{abstract}
The societal impact of rumor spreading is becoming increasingly severe; yet, current research remains relatively one-sided, typically focusing
on either rumor propagation or rumor control while neglecting the confrontational and dynamically evolving relationship between them. To
address this gap, we propose a novel confrontation framework for rumor modeling. We extend the classical Susceptible-Infected-Recovered-Susceptible (SIRS) model into an Ignorant-Spreader-Stifler-Vigilant-Ignorant (SIRQS) framework by introducing a vigilant state and a confrontation mechanism, thereby capturing subtle differences in individual states during rumor
propagation and in their confrontational behavior toward supervisors. At the same time, supervisors patrol the network through random
walks guided by node propagation importance, enabling targeted monitoring of rumor spreaders and individuals with a high risk of spreading rumors. Using a microscopic Markov chain approach, we further characterize heterogeneous node behavior and individual differences,
and couple the propagation and supervision processes to model node-state transition patterns. We conduct simulations on networks with
three different sizes, various topologies, and a real-world network. The results show that the supervision subject, the preference effects associated with the number of supervisors, and the confrontation mechanism are key factors in supervision, and largely determine the effectiveness
of rumor propagation control in the simulations, reflecting the substantial influence of these three mechanisms in real-world spreading
scenarios. Finally, through multiple evaluation indicators, we provide references for determining the optimal number of supervisors.
\end{abstract}

\maketitle
\noindent
\textbf{Current research on rumor dynamics remains relatively one-sided, typically focusing on either rumor propagation or rumor control while neglecting the confrontational and dynamically evolving relationship between them. Here, we address this gap by formulating rumor propagation and supervisory intervention as a coupled confrontation co-evolution process on complex networks and uncover the key factors that govern supervision effectiveness under strategic evasion by rumor spreaders. To model the opposing side of this confrontation, we design an importance-weighted random walk strategy that dynamically directs supervisors toward the most critical nodes and couple both processes via a microscopic Markov chain approach that preserves node-level heterogeneity. This framework reveals that supervision subject, patrol preference, and confrontation intensity constitute three governing factors of rumor control, producing topology-dependent phase transitions and diminishing marginal returns in supervisor deployment. We show that extending the classical SIRS model into an SIRQS framework—with a vigilant state and a proximity-based confrontation mechanism—not only captures heterogeneous individual responses to rumors but also reveals how spreaders’ strategic avoidance of supervisors fundamentally reshapes propagation patterns across different network topologies. Our results establish a unified dynamical picture of rumor–supervision confrontation—from node-level evasion and intervention to network-level containment outcomes—providing actionable guidance for optimal resource allocation in real-world rumor management.}

\noindent\hfil\rule{0.2\textwidth}{0.4pt}\hfil\par
\section{Introduction}
In today's era of information explosion and rapid development of digital media \cite{lazer2018science,festinger1948study}, the rapid expansion of online social networks, and the diversification and acceleration of information flow, rumors can spread more easily and are difficult to eliminate. The widespread spreading of rumors poses serious threats to social stability, public trust, and national democratic governance, among other factors \cite{vosoughi2018spread,horowitz2019disinformation}. Research has shown that rumors are more likely to arise in environments and conditions of high importance and high ambiguity, and the more significant the event and the greater the uncertainty, the stronger the impact of related rumors \cite{difonzo2007rumor}. This phenomenon is particularly evident in public health emergencies and socio-political events. In these situations, false information can heighten public anxiety, undermine institutional credibility, and complicate crisis response efforts \cite{gallotti2020assessing}. Therefore, studying the nature of rumor spreading \cite{wang2024distributed}, as well as controlling and regulating rumor spread \cite{song2019ced}, is an urgent, complex, and multi-dimensional challenge that requires scientifically precise interventions capable of ensuring rational allocation and planning of resources. Today, establishing effective control mechanisms has become a pressing task for experts and scholars in the field of rumor spreading.

As a well-established science of complex systems, network science \cite{Perc2022Socialphysics,boccaletti2006complex,li2017fundamental} provides powerful simplification and modeling tools that open new avenues for theoretical research into capturing realistic phenomena, including rumor spreading and control. A time-varying network \cite{zeng2025bursty,zeng2025complex,feng2024information,li2020evolution,qin2023detecting} is a typical example, which accurately captures the changes in the relationships between individuals during the propagation process through the dynamics of the network structure. Moreover, considerable advances have been made through complex network theory \cite{moreno2004dynamics}, with many studies adopting epidemiological models to describe the dynamics of rumors \cite{daley1964epidemics}. Classical frameworks, such as Susceptible-Infected-Recovered (SIR) \cite{cao2023dynamical}, their extensions, like layered SIRS \cite{zhang2021layered}, and more refined models, including Spreaders-Ignorants-Accepting Stiflers-Unaccepting Stiflers (SIRaRu) \cite{wang2014siraru} and Susceptible-Infected-Hibernator-
Removed (SIHR) \cite{zhao2012sihr}, offer foundational structures for capturing population-level state transitions. In addition, recent studies have further integrated refutation mechanisms inspired by the ``anti-spiral of silence" theory \cite{zhu2018influence}. These approaches classify users into distinct categories, such as Ignorant, Spreader, Debunker, and Resistant, and introduce more refined models such as Susceptible-Infective-Remover-Debunker-Remover (SIRDR) \cite{li2025modeling}. The susceptible-trusted-infectious-recovered (STIR) framework, for instance, differentiates between natural forgetting and immunization through debunking. It also incorporates mechanisms, such as the information exposure attractiveness index to capture finer dynamic features of rumor propagation. Further integrating game theory, some studies have proposed heuristic-enhanced rumor propagation frameworks such as the STIR model, which systematically embeds anti-rumor mechanisms and trust dynamics within a cognitive game-theoretic structure \cite{mou2025information}. Beyond compartmental models, the Microscopic Markov Chain Approach (MMCA) \cite{gomez2010discrete,arenas2023bifurcation} enables granular modeling of state transitions at the node level, accounting for heterogeneity and behavioral variation across individuals. In parallel, research on control strategies often examines platform-led interventions such as edge removal \cite{Perc2024networks,nandi2016methods} and account blocking \cite{yang2023rumor,manouchehri2021temporal}, which aim to disrupt rumor spreading by altering network topology. Complementary studies focus on the spreading dynamics and influencing factors of debunking information \cite{LiXiaoJie2022}. Collectively, these works provide a range of effective strategies and supervisory insights for controlling rumor spreading and limiting its reach by identifying key nodes and critical edges \cite{Shi2023}.

Admittedly, some existing studies have begun to take the dynamics of propagation and supervision into account. However, these studies either focus on expanding the scope of propagation states \cite{he2025efficient} and discussing the increased complexity after integrating game theory \cite{pan2023dynamic}, or directly introduce an anti-rumor mechanism \cite{xiao2020rumor}. Only few of them treat propagation and supervision as two opposing groups that can operate with their own independent mechanisms \cite{LiLulu2020}, which undermines the accurate consideration of the authenticity and dynamics of their confrontation process. Moreover, prevailing supervision strategies, largely built upon oversimplified assumptions, prove inadequate in real-world contexts featuring complex network structures and dynamically evolving scenarios. To fill this research gap, we propose a novel confrontation rumor modeling framework to characterize the dynamic competitive evolutionary process between rumor spreaders and supervisors. For a more precise distinction of population states in the context of rumor propagation and to illustrate how spreaders attempt to bypass supervision to achieve larger-scale rumor spreading, we introduce the rumor-suspecting state Q \cite{li2025rumor} as a vigilant state and a supervision confrontation mechanism into the rumor spreading process, expanding the traditional SIRS framework. Additionally, we model supervisors' patrolling as a random walk \cite{noh2004random} driven by the propagation importance of the nodes, capturing the behavior of supervisors patrolling based on both blocking rumors and tracking spreaders. Furthermore, supervision effectiveness dynamically adjusts according to the state of target nodes, reflecting the varying degrees of supervision effect depending on the influence of rumors on the monitored objects. Then, by using MMCA, we couple the behaviors of the two primary agents to describe node state transition patterns, highlighting node heterogeneity and behavioral differences. Finally, we further simulate various parameters to identify key supervision factors, quantify their impact, and apply our model to real-world networks to demonstrate the relevance of our research to real-life rumor propagation.

The remainder of this paper is structured as follows: Section \uppercase\expandafter{\romannumeral2} details the modeling of the rumor spreading process using our extended model. Section \uppercase\expandafter{\romannumeral3} presents the modeling of supervision strategies and their mobility within the network. Section \uppercase\expandafter{\romannumeral4} provides extensive simulations under various parameter settings to validate the model and explore system dynamics. Section \uppercase\expandafter{\romannumeral5} concludes the paper, summarizing key findings, discussing implications for adaptive rumor control, and outlining directions for future research.

\section{Rumor Spreading MODEL}
Existing studies typically use epidemic models to approximate the modeling of rumors, with scholars categorizing the states of nodes in a rumor network into three types---ignorant, spreader, and stifler \cite{sattenspiel2000epidemic,moreno2004dynamics}---which correspond to the SIR model \cite{cheng2013epidemic,cao2023dynamical,dong2018sis}. Therefore, based on rumor propagation under the SIRS model, we expand it by introducing a vigilant state Q, which is mentioned in literature \cite{li2025rumor}, to represent individuals who remain skeptical and are not easily persuaded after encountering a rumor \cite{azmibehavioral}. We use state Q to reflect different extents of susceptibility among individuals and to exhibit heterogeneous supervision intensity in our subsequent supervision mechanism, determined by target states. Additionally, to simplify our model, the network we study is a closed weighted undirected network, without considering population movement or birth-death processes.

In this section, we will explain the core components of the SIRQS model. We first define the state space and transition rules, then introduce the antagonistic mechanism of rumor spreaders against supervisors.

\subsection{SIRQS Model with Vigilant State}
We now detail the state definitions and transition rules in the SIRQS model. The state space comprises four distinct epidemic states:
\begin{enumerate}
    \item S (Ignorant): Individuals who are unaware of the rumor. They may be infected with rumors through contact with rumor neighbors with probability $\beta$, or be subject to supervision.
    \item Q (Vigilant): Individuals with elevated awareness, exhibiting increased skepticism toward rumors. They possess a reduced probability $\theta\beta$ of being infected with rumors through contact with rumor neighbors and may revert to state S with probability $\gamma_1$, where $\theta\in(0,1)$ represents the vigilant extent of Q state.
    \item I (Spreader): Individuals who actively spread the rumor. They transition to state R with probability $\mu$ due to loss of interest, or may be neutralized via supervision.
    \item R (Stifler):  Individuals who are aware of the rumor but refrain from spreading it, having exited states S or I through supervision. They transition to state Q with probability $\gamma_2$, reflecting sustained moderate vigilance after prolonged rumor-free periods.
\end{enumerate}

We define a stochastic variable $X_i(t)$ to represent the state of node $i$ at time $t$, where $X_i(t) \in \{S, I, R, Q\}$. Let $p_i^X(t)$ denote the probability that node $i$ is state $X$ at time $t$. These probabilities satisfy the normalization condition: $p_i^S(t) + p_i^I(t) + p_i^Q(t) + p_i^R(t) = 1.$
\begin{figure*}[t]
    \centering
    \includegraphics[width=1\linewidth]{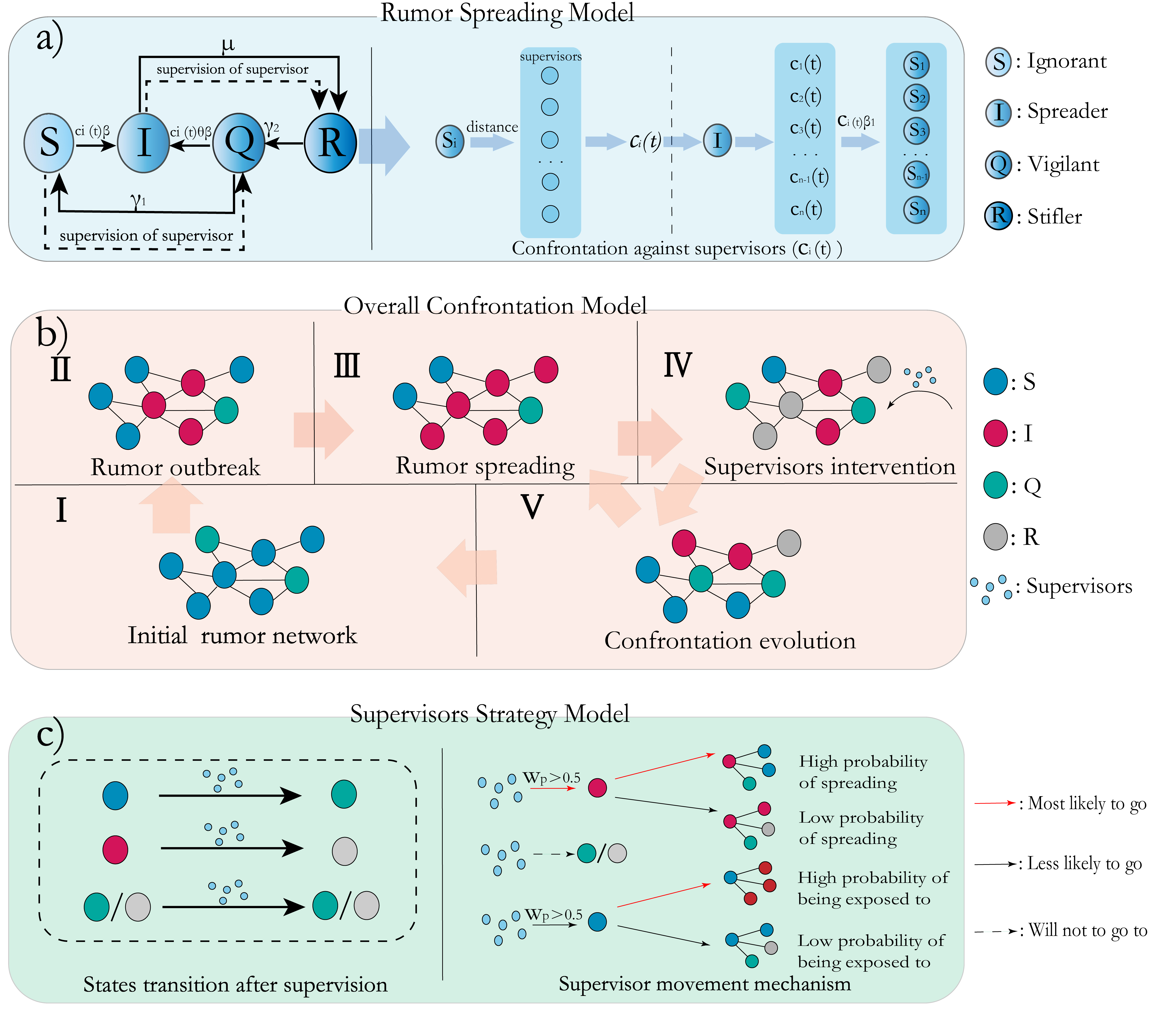}
    \caption{\textbf{Schematic Diagram of the Confrontation Co-evolution Model.} Subfigure (a) illustrates the rumor propagation model. On the left is a state transition diagram containing
four states: S (Ignorant), I (Spreader), Q (Vigilant), and R (Stifler). Arrows indicate the direction and probability of transitions, while dashed arrows represent transitions
triggered by supervision interventions. On the right is the rumor propagation countermeasure mechanism based on the distance between target nodes and supervisors.
Subfigure (b) uses blue, red, green, and gray circles to represent the four states S, I, Q, and R, respectively, showing the overall confrontation evolution process of the system
from an initial rumor-free state to a final cyclic state, with numbers indicating the order of evolution. Under full supervision intervention, the system may eliminate rumors and
return to a rumor-free state. Subfigure (c) shows the supervision mechanism. The left part illustrates the state transitions of nodes under supervision influence. The right
part outlines the movement strategy of supervisors: supervisors move only toward target nodes in states S and I. The parameter $w_p$ (supervisor preference for I nodes)
determines whether they prioritize moving to S nodes for prevention or to I nodes for intervention. Movement preference toward I nodes is based on their rumor propagation
probability, while movement preference toward S nodes is based on their susceptibility to rumors. Red arrows, black arrows, and dashed arrows indicate the directional patrol
preferences of supervisors. Taken together, this schematic illustrates the coupled feedback loop between strategic rumor evasion and targeted supervisory patrol that drives
the system’s co-evolution.}
    \label{fig:schematic}
\end{figure*}
\subsection{Rumor Spreading Mechanism with Confrontation Against Supervision}
In real-world scenarios, rumor spreaders typically adjust their spreading strategies based on actual conditions. For example, they often choose platforms with lighter supervision or groups that receive less attention from supervision authorities for spreading. In this subsection, we integrate this confrontation mechanism into the SIRQS model. We quantify the level of confrontation adjustment factor $c_i(t)$ through the average distance from the node to the supervisors. Rumor spreaders assess the likelihood of the target being supervised through the confrontation adjustment factor, and they will bypass nodes with a higher supervision likelihood, meaning the rumor transmission rate of these nodes will be lower. We also introduce a confrontation coefficient, $\xi$, to quantify the intensity of the spreaders' resistance to supervisors in the current network \cite{pan2022dynamic,ding2025endogenous}. Let $r_n(t)$ denote the position of the $n$-th supervisor at time $t$, and $d_{i \to r_n(t)}$ the hop distance between node $i$ and supervisor $r_n$ at time $t$. The mean distance from node $i$ to all supervisors is then defined as
\begin{equation}
    \overline{d}_{i \to r}(t) = \frac{\sum_{n=1}^{N_r} d_{i \to r_n(t)}}{N_r},
\label{eq:mean_distance}
\end{equation}
where $N_r$ denotes the total number of supervisors. Based on this, the confrontation adjustment factor $c_i(t)$ of node $i$ at time $t$ is defined as a function of the sigmoid type to express the characteristic of the S line\cite{yin2003flexible} of this effect as
\begin{equation}
    c_i(t) =
        \frac{\xi}{1 +e^{-\overline{d}_{i \to r}(t)}},
\label{eq:confrontation_factor}
\end{equation}
where the confrontation parameter $\xi$ modulates how supervisor proximity affects the likelihood of rumor transmission. A higher value of $\xi$ indicates that nodes at varying distances from supervisors exhibit more divergent rumor spreading probabilities, thereby promoting the rumor propagation toward regions farther from supervision coverage.
Then, we normalize it through an upper truncation function as
\begin{equation}
   \hat{c}_i(t)=min\{c_i(t),1\}.
\label{eq:truncation}
\end{equation}

Based on the above, we can define the revised rumor spread probability of each node $i$ at time $t$ as
\begin{equation}
\begin{cases}
    \beta_i^S(t)=\hat{c}_i(t)\beta,\\
    \beta_i^Q(t)=\hat{c}_i(t)\theta \beta.
\end{cases}
\label{eq:revised_spread_prob}
\end{equation}

\section{Rumor Supervision Model}
In this section, we model supervisors as mobile agents performing discrete-time random walks within the network. These supervisors represent real-world entities such as social media moderators. Finally, we integrate this supervision model with the rumor propagation framework using MMCA, thereby establishing a comprehensive confrontation dynamic co-evolution model.

\subsection{Supervised Intervention Based on Spreading Importance}
We first give the effect of supervision, as when a supervisor visits a node, it executes state-specific interventions according to the following rules:
\begin{enumerate}
    \item Target nodes in state S will transition to state Q.
    \item Target nodes in state I will transition to state R.
    \item Target nodes in states Q or R remain unchanged.
\end{enumerate}

Then, for the patrol mechanism, we combine supervisors' random walks with the propagation importance of target nodes, meaning supervisors are more likely to move toward individuals who play a more crucial role in the spread of rumors for supervision control. The transition probabilities of each supervisor are guided by a spreading importance metric $\Psi_j$. This quantity measures the potential contribution of each neighbor node to rumor propagation, incorporating both preventive and interventive considerations through a preference parameter $w_p \in [0,1]$. For a supervisor located at node $i$, the spreading importance of neighboring node $j$ is calculated as:
\begin{equation}
\Psi_j(t) \!=\!
\begin{cases}
\displaystyle
(1 \!-\! W_p) \!\cdot\! \sum_{k \in \mathcal{N}(j)} \beta_j^S(t) \, p_k^I(t) \, a_{jk},
& \!\!X_j(t) \!=\! S, \\[10pt]
\displaystyle
W_p \!\cdot\! \sum_{k \in \mathcal{N}(j)} \!\Bigl[ \beta_j^S(t) p_k^S(t) \!+\! \beta_j^Q(t) p_k^Q(t) \Bigr] a_{jk},
& \!\!X_j(t) \!=\! I,
\end{cases}
\label{eq:spreading_importance}
\end{equation}
where $a_{jk}$ is an element of the adjacency matrix $A$ and represents the connection relationship between nodes $j$ and $k$. Meanwhile, the preference parameter $W_p$ governs the strategic emphasis: higher values prioritize reducing active spreading (nodes in state I), while lower values emphasize protecting ignorant individuals (nodes in state S) \cite{tong2023deterministic}.

We argue that the patrol of supervisors satisfies the Markov property because the transition probability of supervisor $k$ from node $i$ to $j$ at moment $t$ depends solely on its position and the states of its target nodes and neighbors at the last moment $t-1$. Therefore, we define a three-dimensional time-varying reach matrix $W(t)$ to capture the probability of supervisors walking among nodes, whose elements are:
\begin{equation}
    w^k_{i  j}(t) = \frac{\delta_i^k(t-1)\Psi_j(t)}{\sum_{w \in \mathcal{N}(i)} \Psi_w(t)},
\label{eq:reach_matrix}
\end{equation}
where $\delta_i^k(t)\in\{0,1\}$ represents the supervisor $k$ at or not at node $i$ at moment $t-1$, respectively. This matrix directs supervisors toward nodes exhibiting higher spreading potential and importance based on their current state and network context.

Based on Eq.~(\ref{eq:reach_matrix}), the probability of node $i$ being supervised depends on the collective presence of all supervisors as:
\begin{equation}
\rho^{S}_i(t) =\rho^{I}_i(t)= 1 - \prod_{k=1}^{N_r} \left[1 - \sum_{j=1}^{\mathcal{N}(i)} w^k_{ji}(t)\right],
\label{eq:supervision_prob}
\end{equation}
where $\rho^{S}_i(t) $ and $\rho^{I}_i(t)$ represents the probability of node $i$ being supervised in states S and I, respectively.

\subsection{Integrated Spreading-Supervision Dynamics}
To further illustrate the confrontation interaction between our rumor propagation and supervision models, and taking into account the significant heterogeneity among nodes in our model, in this section, we use the MMCA \cite{gomez2010discrete,wu2011epidemic} to couple the mechanisms and models of the two main processes. The complete system integrates these two processes through a unified set of equations that accounts for the stochastic movement of the supervisors and their impact on the evolution of node states, with both processes evolving in discrete time steps. Let us denote $a_{ij}$ the adjacency matrix that supports the SIRQS process. Then the transition probabilities for node $i$ in state S and Q being informed by any neighbors, respectively, $q_i^S(t)$ and $q_i^Q(t)$ are
\begin{equation}
\begin{cases}
q_i^S(t) = 1 - \displaystyle\prod_{j=1}^{n} \left[1 - a_{ji} \beta_i^S(t)  p_j^I(t) \right], \\[8pt]
q_i^Q(t) = 1 - \displaystyle\prod_{j=1}^{n} \left[1 - a_{ji} \beta_i^Q(t) p_j^I(t) \right].
\end{cases}
\label{eq:informed_prob}
\end{equation}
Using Eqs.~(\ref{eq:supervision_prob}) and~(\ref{eq:informed_prob}), the complete nodal state evolution under combined spreading and supervision dynamics is given by

\begin{equation}
\begin{cases}
p_i^S(t+1) = [1 - q_i^S(t)] \cdot [1 - \rho_i^S(t)] \cdot p_i^S(t) + \gamma_1 \cdot p_i^Q(t) \cdot \\ [1 - q_i^Q(t)],  \\
p_i^Q(t+1) = [1 - q_i^Q(t)] \cdot (1 - \gamma_1) \cdot p_i^Q(t) + \gamma_2 \cdot p_i^R(t) +\\ \rho_i^S(t) \cdot p_i^S(t), \\
p_i^I(t+1) = q_i^S(t) \cdot [1 - \rho_i^S(t)] \cdot p_i^S(t) + q_i^Q(t) \cdot p_i^Q(t) +\\(1 - \mu) \cdot [1 - \rho_i^I(t)] \cdot p_i^I(t), \\
p_i^R(t+1) = (1 - \gamma_2) \cdot p_i^R(t) + \bigl[\mu \cdot [1 - \rho_i^I(t)] + \rho_i^I(t)\bigr] \cdot \\ p_i^I(t).
\end{cases}
\label{eq:state_evolution}
\end{equation}

At each time step, the node state is updated according to Eq.~(\ref{eq:state_evolution}), while the supervisor's position evolves according to Eq.~(\ref{eq:reach_matrix}); The two processes interact through the intervention probabilities $\rho_i^S(t)$, $\rho_i^I(t)$, and the confrontation revised rumor spread probabilities $q_i^S(t)$ and $q_i^Q(t)$.

To show our model more clearly and directly, a schematic diagram of the rumor spreading process is provided in Fig.~\ref{fig:schematic} for visual illustration. We presented it from three perspectives: rumor spreading, supervision, and the dynamic confrontation co-evolution of the two behaviors.

To present our model more clearly and easily, we have listed the relevant symbolic explanations in Tab.~\ref{tab:symbol_three_line}.

\begin{table}[h]
\caption{\label{tab:symbol_three_line}Symbols and Plain-Language Definitions for the Main Model Parameters. This table provides a unified reference for all symbols used throughout the model description, state transition equations, and simulation analyses.}
\begin{ruledtabular}
\begin{tabular}{l p{0.4\textwidth}}
\textbf{Symbol} & \multicolumn{1}{c}{\textbf{Explanation}} \\ 
\hline
$\beta$        & Rumor spreading rate \\
$\theta$       & The vigilant extent of state $Q$ \\
$\gamma_1$     & The rumor disbelief lose rate from state $Q$ to $S$ \\
$\gamma_2$     & The rumor disbelief lose rate from state $R$ to $Q$ \\
$\mu$          & The rate of state $I$ loses interest turn to state $R$ \\
$\xi$          & The confrontation parameter indicating the confrontation extent of nodes in state $I$ \\
$N_r$          & The number of supervisors \\
$\mathcal{N}(i)$ & The set of neighbors of node $i$ \\
$W_p$          & The preference parameter indicating the extent of the supervisor's patrol preference to node in state $I$ \\
\end{tabular}
\end{ruledtabular}
\end{table}
\subsection{Steady-State Analysis and Spreading Threshold}
In this section, we determine the spreading threshold through linear stability analysis around the rumor-free equilibrium (RFE) \cite{zhong2023rumor,musa2019mathematical}, where $p_i^I = 0$ for all nodes $i$. At RFE, the nodal state distribution satisfies the equilibrium conditions derived from Eq.~(\ref{eq:state_evolution}):
\begin{equation}
\begin{cases}
p_i^S(t)=p_j^S(t)=p^S_i = (1 - \rho^*) \cdot p^S_i + \gamma_1 \cdot p^Q_i,\\
p_i^Q(t)=p_j^Q(t)=p^Q_i = (1 - \gamma_1) \cdot p^Q_i + \gamma_2 \cdot p^R_i + \rho^*_i \cdot p^S_i, \\
p_i^I(t)=p_j^I(t)=p_i^I = 0,\\
p_i^R(t)=p_j^R(t)=p^R_i = (1 - \gamma_2) \cdot p^R_i.
\end{cases}
\label{eq:rfe_conditions}
\end{equation}

The uniform value $\rho^*$ represents the steady-state values of $\rho^{S}_i(t)$ and $\rho^{I}_i(t)$ resulting from a minimal perturbation of rumor individuals, indicating a stable supervision capability.

Solving this system yields the RFE distribution:
\begin{equation}
p^{S}_i = \frac{\gamma_1}{\gamma_1 + \rho^*}, \quad p^{Q}_i = \frac{\rho^*}{\gamma_1 + \rho^*}, \quad p^{R}_i = 0.
\label{eq:rfe_distribution}
\end{equation}

Linearizing the system around RFE with small perturbations $\epsilon_i = p_i^I$ yields the evolution equation. Considering only first-order terms in $\epsilon_i$, the rumor dynamics simplify to
\begin{equation}
\epsilon_i = q_i^S (1 - \rho^*) p_i^S + q_i^Q p_i^Q + (1 - \mu)(1 - \rho^*) \epsilon_i,
\label{eq:linearized_evolution}
\end{equation}
where $q_i^S$ and $q_i^Q$ are $q_i^S(t)$ and $q_i^Q(t)$ in Eq.~(\ref{eq:informed_prob}) linearized as
\begin{equation}
\begin{cases}
    q_i^S \approx \sum_j a_{ji} \beta_i^S(t) \epsilon_j=\sum_j a_{ji} \hat{c}_i \beta \epsilon_j, \\
    q_i^Q \approx \sum_j a_{ji} \beta_i^Q(t) \epsilon_j=\sum_j a_{ji} \hat{c}_i\theta \beta \epsilon_j.
\end{cases}
\label{eq:linearized_q}
\end{equation}
Substituting into the $\epsilon_i$ equation:
\begin{equation}
\epsilon_i = \hat{c}_i \beta \left[ (1 - \rho^*) p^S_i + \theta p^Q_i \right] \sum_j a_{ji} \epsilon_j + (1 - \mu)(1 - \rho^*) \epsilon_i.
\label{eq:substituted_epsilon}
\end{equation}

In matrix form, this becomes:
\begin{equation}
\boldsymbol{\epsilon} = \mathbf{M} \boldsymbol{\epsilon},
\label{eq:matrix_form}
\end{equation}
where the matrix $\mathbf{M}$ has elements:
\begin{equation}
M_{ij} = \hat{c}_i \beta \left[ (1\! -\! \rho^*) p^S_i + \theta p^Q_i \right] a_{ji} + (1\! -\! \mu)(1\! -\! \rho^*) \delta_{ij},
\label{eq:matrix_elements}
\end{equation}
where $\delta_{ij}$ denotes the Kronecker delta function. Combining the steady-state formula Eq.~(\ref{eq:rfe_distribution}), we can obtain the matrix $M$ as
\begin{equation}
\mathbf{M} = \beta \frac{ (1\! -\! \rho^*) \gamma_1 + \theta \rho^* }{ \gamma_1 + \rho^* } \text{diag}(\hat{c}_i) \mathbf{A}^T + (1\! -\! \mu)(1\! -\! \rho^*) \mathbf{I}.
\label{eq:matrix_M}
\end{equation}

The critical spreading threshold occurs when the spectral radius satisfies $\rho(\mathbf{M}) = 1$, marking the transition between rumor extinction and persistence. For homogeneous networks, we can derive an explicit critical transmission rate by noting that the largest eigenvalue of the matrix $\mathbf{M}$ corresponds to
\begin{equation}
\Lambda_{\text{max}}(\mathbf{M}) = \beta \frac{ (1\! -\! \rho^*) \gamma_1 + \theta \rho^* }{ \gamma_1 + \rho^* } \Lambda_{\text{max}}(L) + (1\! -\! \mu)(1\! -\! \rho^*),
\label{eq:max_eigenvalue}
\end{equation}
where matrix $L=\text{diag}(\hat{c}_i)\cdot \mathbf{A}^T$. Setting $\rho(\mathbf{M}) = 1$ and substituting the RFE solutions for $p^S$ and $p^Q$ yields the following:
\begin{equation}
\beta_c = \frac{  \left[ \mu + \rho^* (1 - \mu) \right] (\gamma_1 + \rho^*) }{ \Lambda_{\text{max}}(L) \left[ (1 - \rho^*) \gamma_1 + \theta \rho^* \right] }.
\label{eq:threshold}
\end{equation}

This threshold is regulated by multiple factors: increasing the intervention probability $\rho^*$ or decreasing $\mu$ and $\theta$ can raise the threshold, thereby suppressing rumor spread; whereas the sparser the network (the larger $\Lambda_{\text{max}}$), the lower the threshold, making rumors more likely to spread. It is worth noting that the opposition ability of the spreaders ($\text{diag}(\hat{c}_i)$) has the opposite effect---the stronger the resistance, the higher the threshold, which is more likely to trigger a rumor outbreak. Finally, we find that when there is no regulation ($\rho^* = 0$), the model threshold degenerates into the classical SIS form $\beta_c = \frac{\mu}{\Lambda_{\text{max}}(L)}$.

\section{SIMULATION}
In this section, we conduct a series of simulations on the aforementioned model. These simulations aim to evaluate the performance of the proposed integrated confrontation model in rumor propagation scenarios and in the Caltech Facebook real-world network \cite{nr}, as well as to analyze the sensitivity of its core mechanisms. Preliminary configuration references for supervisors are provided based on several evaluation metrics. Most of the simulations are carried out on a Barab\'asi--Albert (BA) network \cite{barabasi1999emergence} containing 1,000 nodes. Additionally, to explore the impact of network topology, further comparative analyses are conducted among BA, Watts--Strogatz (WS)\cite{watts1998collective}, and Erd\H{o}s--R\'enyi (ER) network \cite{erdds1959random} models. To ensure the robustness and statistical reliability of the results, each simulation setup is run independently 20 times, and the final results are reported as the mean of these simulations.

\begin{figure*}[t]
    \centering
    \includegraphics[width=\linewidth]{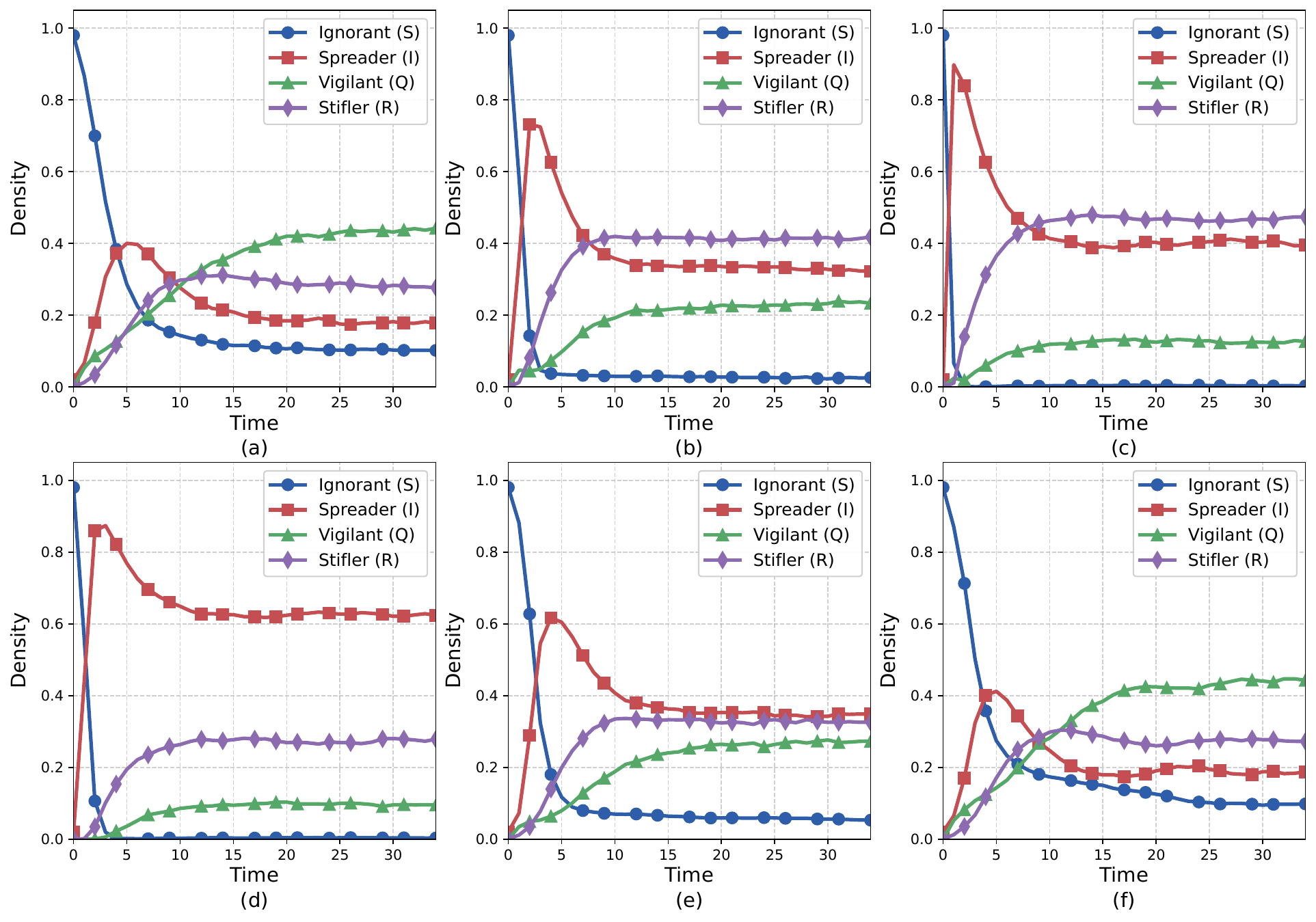}
    \caption{\textbf{The Iterative Spread of Rumors Under Different Rumor Transmission Rates and Numbers of Supervisors.} The horizontal axis represents the time steps of rumor spreading, and the vertical axis represents the density of nodes in each state. The four solid lines correspond to four types of node states: blue represents S (Ignorant), magenta represents I (Spreader), orange represents Q (Vigilant), and red represents R (Stifler). Subfigures (a)-(c) show simulation results under different rumor transmission rates $\beta$: from left to right, $\beta=0.15$, $\beta=0.3$, $\beta=0.75$, illustrating the impact of transmission strength on state evolution with the same vigilance factor $\theta=1/3$. Subfigures (d)-(f) show simulation results for different numbers of supervisors $N_r$; all subfigures use $\beta=0.15$ and $\theta=1/3$, corresponding, from left to right, to $N_r=0$, $30$, and $50$ supervisors, to demonstrate the effect of resource allocation on rumor suppression. The comparison reveals that higher transmission rates accelerate outbreaks and increase the endemic level of spreaders, whereas deploying more supervisors suppresses the peak infection and promotes the accumulation of vigilant individuals.}
    \label{fig:iterative_spread}
\end{figure*}

\subsection{Impact of Rumor Transmission Rates and Numbers of Supervisors}
The rumor spreading rate of S nodes and the number of supervisors are key factors affecting rumor propagation and supervision effectiveness. This section first explores the roles of these two factors through iterative evolutionary simulations of the model, as shown in Fig.~\ref{fig:iterative_spread}.

By observing all the subfigures in Fig.~\ref{fig:iterative_spread}, we can see that the steady-state density of S nodes remains at a relatively low level; after the system stabilizes, the network is mainly composed of Q, R, and I nodes. This phenomenon further suggests that the S is difficult to maintain for an extended period under such a competitive environment. Moreover, further observation indicates that the density of I nodes initially rises sharply to a peak and then declines to a stable level. This phenomenon is mainly caused by the fact that the initial positions of supervisors are not I nodes, so they must spend time passing through S nodes first. Secondly, during this period, the I nodes actively spread rumors while strategically avoiding supervisors. When a large number of supervisors eventually reach the I nodes, the dynamic confrontation between propagation and supervision eventually stabilizes.

Comparing Figs.~\ref{fig:iterative_spread}(a)--(c) with different propagation rates, we see that the higher the rumor propagation rate, the faster the density of I nodes increases, the steeper the rising curve, and the higher the peak. Moreover, higher propagation rates also result in higher steady-state densities of I and R nodes, and lower steady-state densities of S and Q nodes. This is because under a high propagation rate, S-state nodes are quickly exposed to rumors from I nodes. On the one hand, supervisors cannot match the speed of rumor propagation, and on the other hand, a large number of nodes remain in the I; therefore, supervisors primarily target the rumor nodes.

Finally, Figs.~\ref{fig:iterative_spread}(d)--(f) with different numbers of supervisors illustrate that the more supervisors there are, the more inhibited the growth of I node density becomes, with a lower peak and gentler upward trend. In addition, deploying more supervisors reduces the steady-state density of I nodes by improving the efficiency and timeliness of rumor supervision and increases the steady-state density of Q-state nodes.

\begin{figure*}[t]
    \centering
    \includegraphics[width=\linewidth]{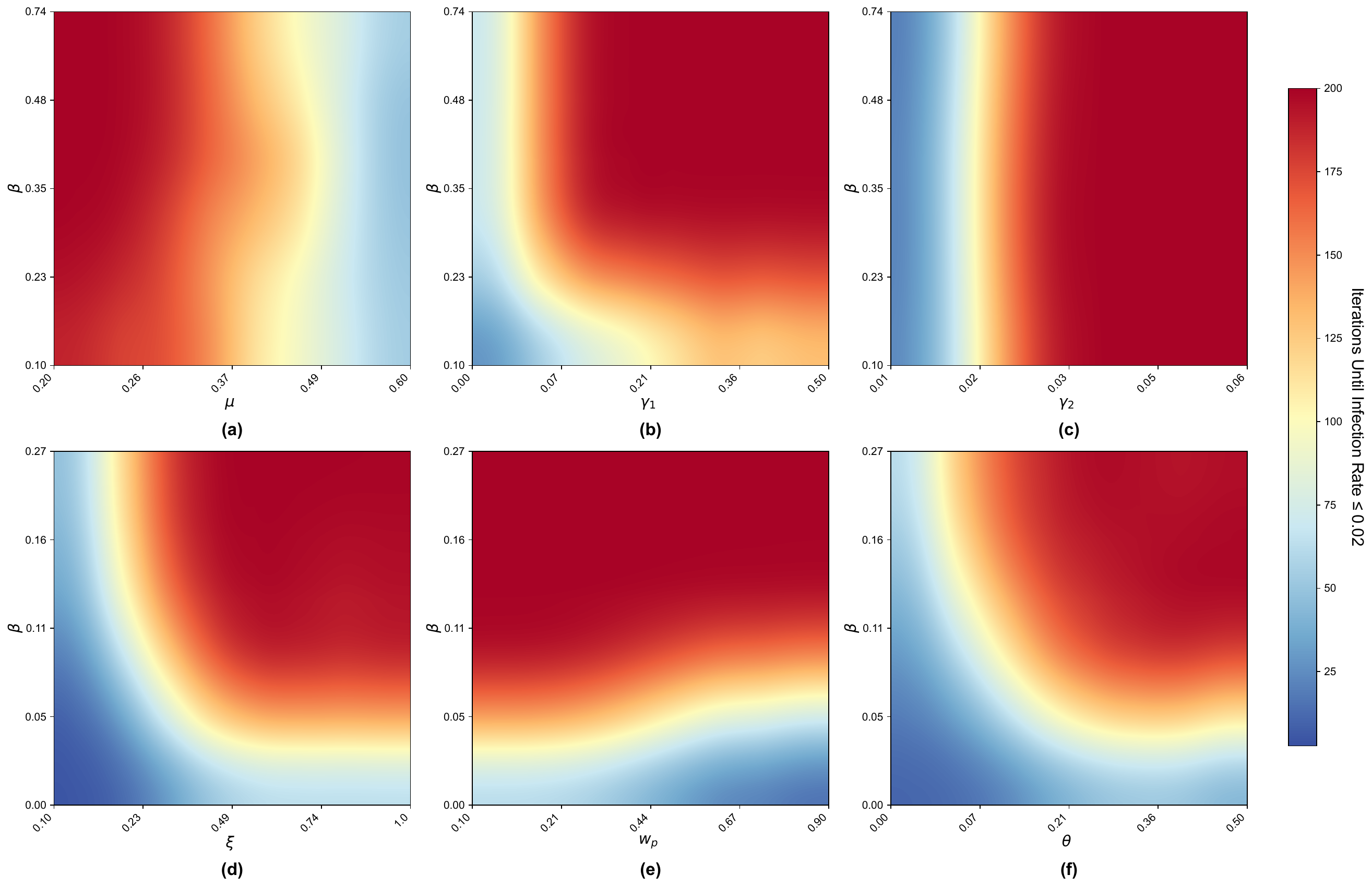}
    \caption{\textbf{Impact of Parameter Variations on Rumor Spreading and Control.} The vertical axis in all six heatmaps represents the rumor transmission rate $\beta$ of S nodes. In the top row, the horizontal axes correspond to the stifling rate $\mu$, the reversion rate $\gamma_1$ (Q $\rightarrow$ S), and the reversion rate $\gamma_2$ (R $\rightarrow$ Q), respectively. In the bottom row, they represent the parameters more relevant to the spreading confrontation and supervision mechanisms: the confrontation intensity $\xi$, the supervisor preference $w_p$ (preference for I-nodes), and the vigilance factor $\theta$. Each heatmap shows the number of time steps required to reduce the density of rumor nodes below 0.02. The color gradient, from blue to red, indicates the duration needed for rumor containment, ranging from 0 to over 200 steps. The heatmaps highlight that while $\gamma_1$ and $\gamma_2$ exhibit sharp phase boundaries, the confrontation parameters $\xi$ and $\theta$ systematically delay containment, especially at high $\beta$, whereas the supervisor preference $w_p$ yields more modest improvements when oriented toward I-nodes ($w_p>0.5$).}
    \label{fig:parameter_impact}
\end{figure*}

\subsection{Impact of Spreading Mechanism and Supervision Preference}
This subsection analyzes the impact of key model parameters on rumor propagation. We focus on the roles of rumor propagation parameters $\mu$, $\gamma_1$, $\gamma_2$, and factors related to the supervisors' movements, with the results shown in Fig.~\ref{fig:parameter_impact}.

Overall, in Figs.~\ref{fig:parameter_impact}(d)--(f), the effect of $\beta$ is particularly prominent. An increase in rumor propagation intensity heightens the difficulty of supervision, and a larger value significantly delays the time for the rumor density to drop below 0.02. In contrast, in Figs.~\ref{fig:parameter_impact}(a)--(c), dominated by rumor propagation parameters, the effect of $\beta$ is masked by the propagation dynamics.

As far as the parameters related to rumor propagation are concerned, Fig.~\ref{fig:parameter_impact}(a) shows that the increase of $\mu$ will lead to a decrease in supervision efficiency, and under a larger $\mu$, nodes will spontaneously abandon rumors and form synergy with supervision, thereby improving the control effect. In Fig.~\ref{fig:parameter_impact}(b), when the high $\beta$ with $\gamma_1$ is around 0.05 and the low $\beta$ is around 0.28, there are two phase transition boundaries in the result of system supervision efficiency, because the increase of $\gamma_1$ promotes the $Q\rightarrow S$ transformation and indirectly increases the possibility of rumor spreading, and the $\beta$ value affects the basic density of the rumor nodes in the network, thus indirectly affecting the size of this phase transition value. In Fig.~\ref{fig:parameter_impact}(c), the effect of $\gamma_2$ is more significant, and a clear phase transition boundary is visible around 0.02. Since there are many I nodes in the system and the density of R nodes is higher than that of the Q nodes, the $\gamma_2$ that regulates the $R\rightarrow Q$ process has a greater impact on the overall system.

We next look at the effects of directly confronting related parameters. Figs.~\ref{fig:parameter_impact}(d) and~(f) show that the effects of $\xi$ and $\theta$ are highly similar: both increases will increase the difficulty of rumor control, especially when $\beta$ is high. We believe that the increase of $\xi$ and the increase of the ability of communicators to evade supervision lead to a rapid increase in the difficulty of supervision, while $\theta$ increases the propagation tendency of Q nodes, indirectly reduces the significance of supervision towards S nodes, and thus weakens the supervision effectiveness.

Finally, the influence of $w_p$ in Fig.~\ref{fig:parameter_impact}(e) is relatively mild as it increases, but as it increases to 0.5, its effect gradually becomes prominent: $w_p$ is greater than 0.5, and the supervision of I-state nodes by supervisors is more accurate, and targeted supervision improves the supervision effectiveness.

\begin{figure*}[ht]
    \centering
    \includegraphics[width=\linewidth]{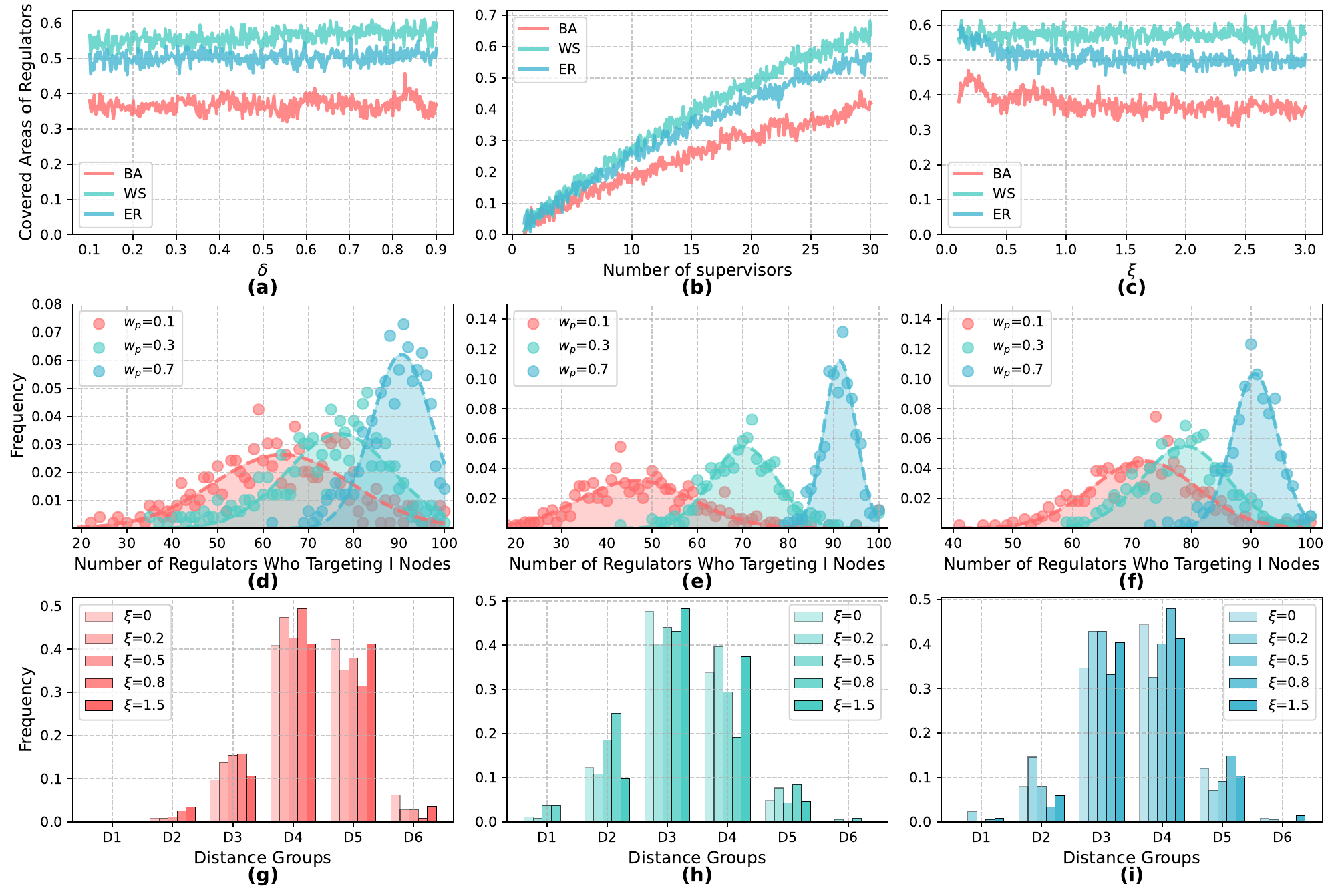}
    \caption{\textbf{Impact of Multiple Factors on the Effectiveness of Supervision.} The first row of subfigures shows the coverage of supervisors, as represented by the steady-state number of independent nodes supervised by any supervisor. Subfigures (a) to (c) show the influence of the rumor propagation rate $\beta$, the number of supervisors $N_r$, and the confrontation parameter $\xi$ on the coverage of supervisors, respectively, where the pink, blue, and green lines correspond to the results of three networks: BA (Barabási–Albert), WS (Watts–Strogatz), and ER (Erdős–Rényi), respectively. In the second and third rows, the simulations' results of the BA, WS, and ER networks are displayed from the first column to the third column. The second row of the subgraph shows the frequency of supervisors moving to I nodes under different supervisor preferences $w_p$ (preference for I-nodes), with red, green, and blue scatters representing $w_p = 0.1$, $0.3$, and $0.7$, respectively. At the same time, we use red, green, and blue dashed lines to represent the fitted curves of their frequency distributions, respectively. The third row of subgraphs shows the average distance from each I node to the supervisor: the distance values are evenly divided into six groups from D1 (closest) to D6 (farthest). The bar chart from light to dark represents the number of nodes in the corresponding range to the average distance to the supervisor under confrontation parameters $\xi = 0$, $0.2$, $0.5$, $0.8$, and $1.5$, respectively. Overall, these results demonstrate that while network topology primarily governs coverage (WS being most explorable, BA least), the supervisor preference $w_p$ enhances the targeting consistency of spreaders, and higher confrontation intensity $\xi$ causes spreaders to cluster farther from supervisors, an effect most pronounced in scale-free networks.}
    \label{fig:multiple_factors}
\end{figure*}
\subsection{Impact of Multiple Factors on the Effectiveness of Supervision}

In the supervisor model, both the supervisor's mobility preferences and the rumor spreader's confrontation behavior have a crucial impact on the scope and effectiveness of regulation. In the simulations of this subsection, we examined results for different $w_p$ and $\xi$, and considered the number of supervisors as a factor in the analysis. In addition, we argue that the network topology is a key factor affecting the simulation results as well. To this end, we conducted simulations on three different types of networks: the BA network, the WS network, and the ER network. The related results are shown in Fig.~\ref{fig:multiple_factors}.

\begin{figure*}[htb]
    \centering
    \includegraphics[width=\linewidth]{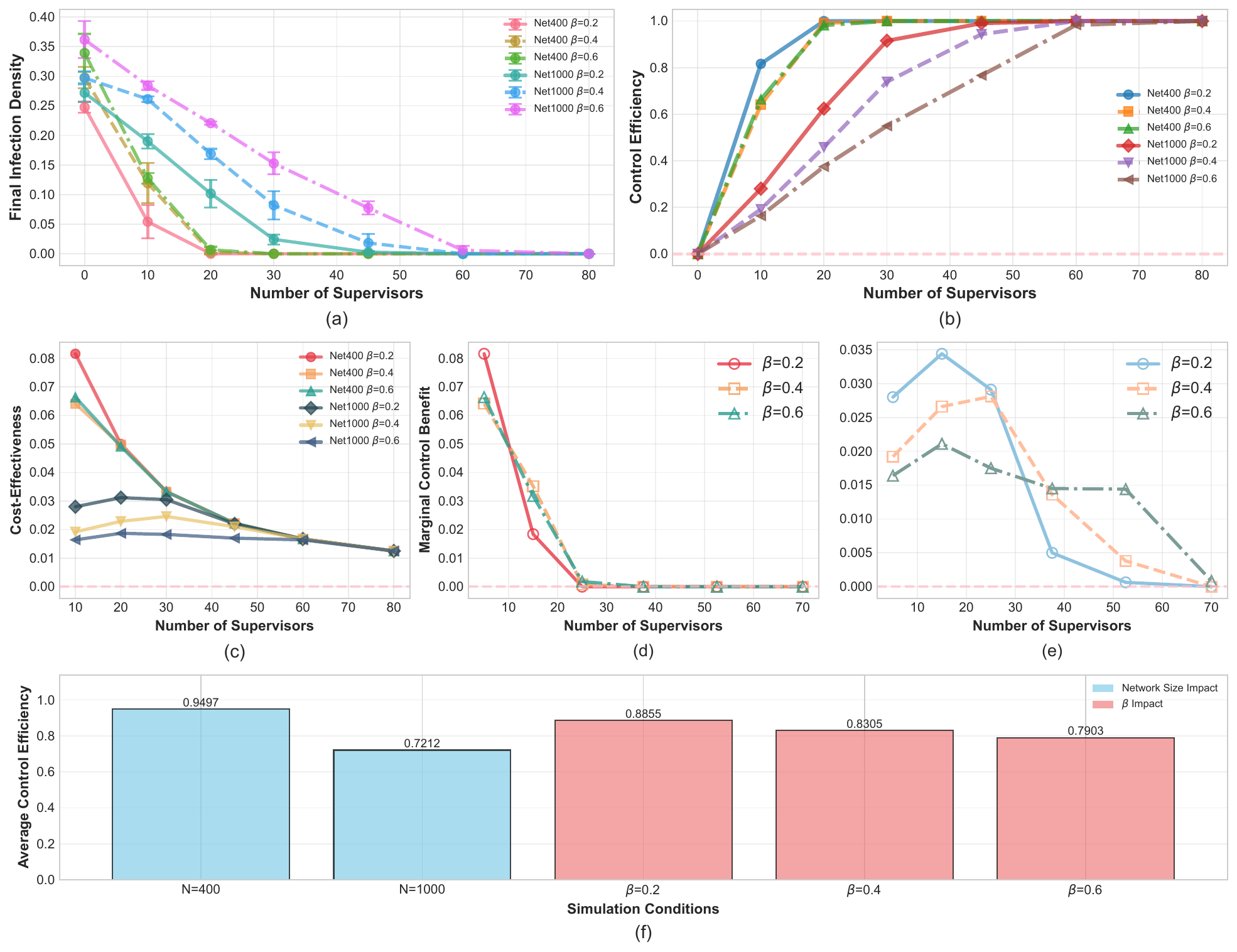}
    \caption{\textbf{Comprehensive Evaluation of Supervision Effectiveness and Optimal Supervisor Allocation.} All subfigures discuss their respective evaluation indicators under different combinations of network scale ($N=400,1000$) and rumor transmission rates ($\beta = 0.2, 0.4, 0.6$). Subfigure a illustrates how the final rumor density varies. Subfigure b presents the corresponding control efficiency $C_E$ (relative reduction in spreader density). In subfigure c, the cost-effectiveness $C_e = C_E / N_r$ of supervision intervention is shown. Subfigures d and e display the marginal control benefit $M$ (incremental gain per additional supervisor) under 400-node and 1000-node networks, respectively, plotted against the midpoint of supervisor number intervals. Finally, subfigure f compares the average control efficiency across simulation conditions, with blue and pink bars distinguishing the effects of network size and the rumor transmission rates, respectively. The evaluation reveals that smaller networks and lower transmission rates achieve higher efficiency with fewer supervisors; cost-effectiveness declines with additional supervisors, indicating diminishing returns; and marginal benefit peaks at low supervisor counts, underscoring the importance of early intervention.}
    \label{fig:comprehensive_eval}
\end{figure*}

\begin{table*}[ht]
\caption{\label{tab:network_scale_transmission_performance}Performance Analysis of Supervision Across Network Sizes and Transmission Rates. The table reports, for two network sizes ($N=400$ and $N=1000$) and three rumor transmission rates ($\beta=0.2,0.4,0.6$), the minimal number of supervisors required to achieve full control efficiency (Optimal Ctrl Supervisor Num), the corresponding control efficiency $C_E$, the number of supervisors that maximizes cost-effectiveness (Optimal Cost Supervisor Num), and the achieved cost benefit $C_e$. The results indicate that smaller networks and lower transmission rates permit complete rumor suppression with substantially fewer supervisors and yield markedly higher cost-effectiveness per supervisor.}
\begin{ruledtabular}
  \begin{tabular}{llcccc}
    \textbf{Network} & \textbf{Transmission Rate} & \textbf{Optimal Ctrl Supervisor Num} & \textbf{Control Efficiency} & \textbf{Optimal Cost Supervisor Num} & \textbf{Cost Benefit} \\
    \hline
     & $\beta$=0.2 & 20 & 1.0000 & 10 & 0.0816 \\
    $N=400$ & $\beta$=0.4 & 30 & 1.0000 & 10 & 0.0641 \\
     & $\beta$=0.6 & 30 & 1.0000 & 10 & 0.0664 \\
    \\
     & $\beta$=0.2 & 60 & 1.0000 & 10 & 0.0312 \\
    $N=1000$ & $\beta$=0.4 & 60 & 1.0000 & 10 & 0.0246 \\
     & $\beta$=0.6 & 80 & 1.0000 & 10 & 0.0187 \\
  \end{tabular}
\end{ruledtabular}
\end{table*}

First, we analyze the coverage of supervisors under different network topologies through Figs.~\ref{fig:multiple_factors}(a)--(c). We find that the BA network is the most resistant to supervision, followed by the ER random network, whose limitations on supervision coverage are slightly better than those of the WS network. The reason for this situation is that the ``hub nodes" in the BA network excessively attract supervisors and reduce coverage efficiency. In contrast, the uniform connectivity and shorter average path lengths of the WS network help supervisors explore the entire network more efficiently and quickly. Since the structural characteristics of the ER network lie between the two, the coverage of supervision is correspondingly in a middle range. It is worth noting that neither the supervisors' preference shown in Fig.~\ref{fig:multiple_factors}(a) nor the confrontation behavior of rumor spreaders shown in Fig.~\ref{fig:multiple_factors}(c) has a significant impact on the coverage area of supervisors. We believe this is because these two parameters primarily optimize the supervisors' current, local decision-making rather than having a global, long-term effect. The dominant factors determining the coverage area are still the macrostructural characteristics of the network and the number of supervisors, as shown in Fig.~\ref{fig:multiple_factors}(b).

Thus, to more accurately delineate the roles of preference and confrontation, we conducted targeted simulations in Figs.~\ref{fig:multiple_factors}(d)--(f) and Figs.~\ref{fig:multiple_factors}(g)--(i). In Figs.~\ref{fig:multiple_factors}(d)--(f), we found that the stronger the supervisor's preference for I nodes, the higher the peak in movement frequency, indicating that the supervisor moves toward I nodes more often. Additionally, the scatter variance under high I node preference is smaller. This suggests that within a fixed time step, supervision decisions are more consistent, and the supervisor's movement patterns are more stable. We further examined the combined impact of network topology and supervision preference. The effect of preference is most pronounced in the WS network, moderate in the ER networks, and weakest in the BA network for the same reason as that for the differences in Figs.~\ref{fig:multiple_factors}(a)-(c).

Finally, we analyze the effect on resistance strength $\xi$ through Figs.~\ref{fig:multiple_factors}(g)--(i). It was found that the peak of the long-distance interval (D3-D5) was higher at high $\xi$, while the frequency was more evenly distributed in the low distance interval at low $\xi$. This indicates that at low confrontation intensity, rumor dissemination is less targeted and more general, so the distribution of rumor spreaders is more balanced. However, we noticed that when the difference in confrontation intensity is small, the long-distance propagation frequency under low confrontation intensity will be lower than that of long-distance propagation under high confrontation intensity, because regulators will also conduct targeted search and supervision of rumor propagation, and this pertinence will be further enhanced in high-confrontation propagation scenarios. In addition, we compare the results of three subfigures representing the three networks and find that the aggregation effect of the BA network is more obvious, and the frequency of the D5 interval is much higher than that of the other two networks under high confrontation, which is similar to the network structure analysis of other simulation structures in this section.

\subsection{Validation on the Caltech Facebook Real-World Network}
\begin{figure*}[ht]
    \centering
    \includegraphics[width=\linewidth]{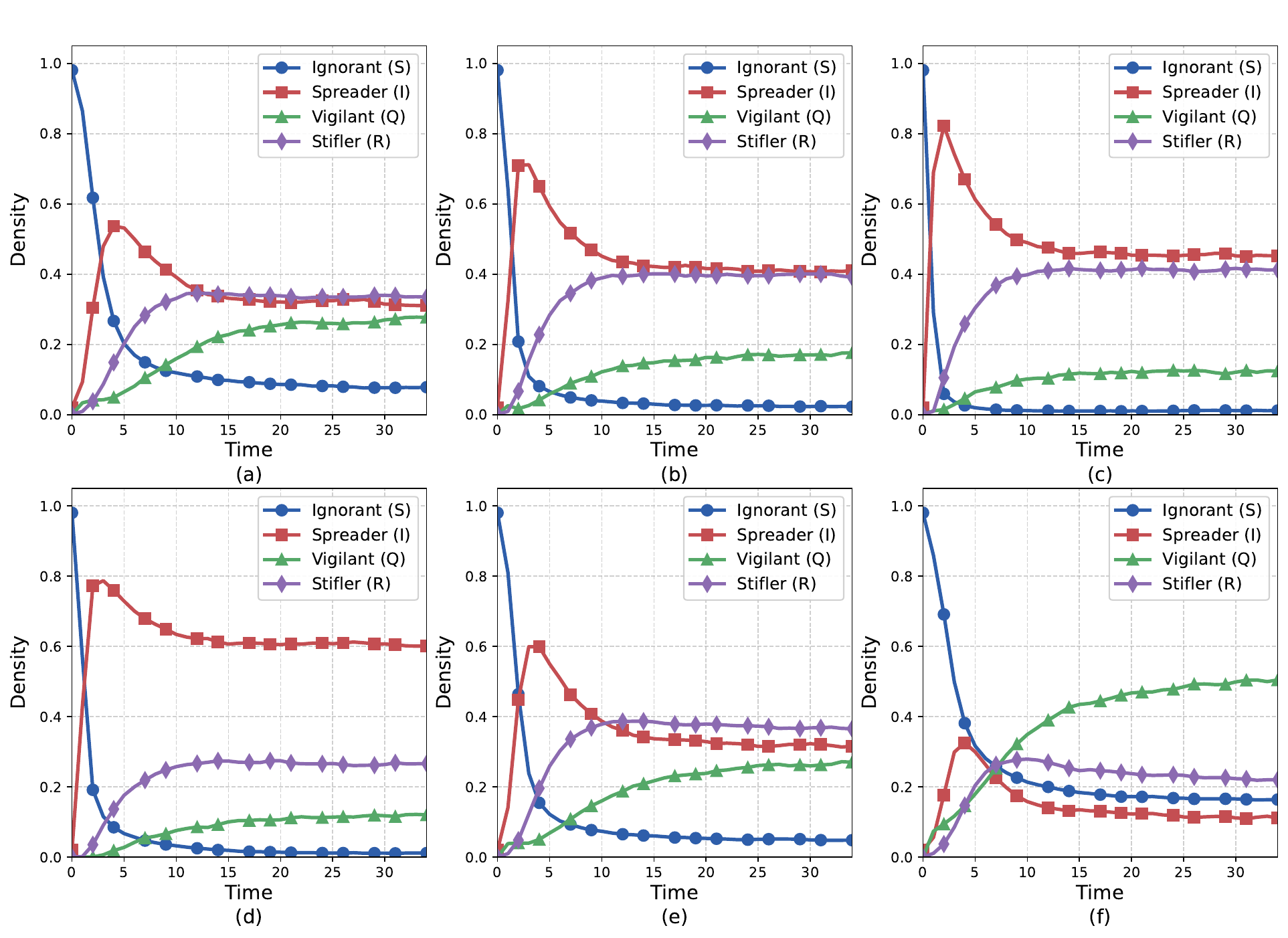}
    \caption{\textbf{The Simulation on the Caltech Facebook Network.} The horizontal axis represents the time steps of rumor spreading, and the vertical axis represents the density of nodes in each state. The four solid lines correspond to four types of node states: blue represents S (Ignorant), magenta represents I (Spreader), orange represents Q (Vigilant), and red represents R (Stifler). Subfigures (a)-(c) show simulation results under different rumor transmission rates $\beta$ (infection probability) with a uniform number of supervisors $N_r = 50$: from left to right, $\beta=0.12$, $\beta=0.25$, and $\beta=0.6$, illustrating the impact of transmission strength on state evolution. Subfigures (d)-(f) show simulation results with different numbers of supervisors $N_r$ (patrolling moderators); all subfigures use $\beta=0.2$, corresponding from left to right to $N_r=0$, $60$, and $120$ supervisors, to demonstrate the effect of resource allocation on rumor suppression. These results on a real-world social network confirm the synthetic-network findings: higher transmission rates amplify outbreaks, while increasing supervisor count curtails spreader peaks and elevates vigilant populations. However, even with extensive supervision, complete eradication remains elusive, reflecting the inherent challenge of rumor control in realistic social topologies.}
    \label{fig:caltech}
\end{figure*}
\subsection{Comprehensive Evaluation of Supervision Effectiveness and Optimal Supervisor Allocation}
To evaluate the effectiveness of our supervisors' supervision strategies, we assess them through four aspects: final rumor density, control efficiency, cost-effectiveness, and marginal control benefit, then provide a supervisor configuration for reference in Fig.~\ref{fig:comprehensive_eval}.
We firstly define the three latter aspects mentioned above:
\begin{enumerate}
    \item Control efficiency($C_E$): $C_E=\frac{p^I_{non-s}-p^I}{p^I_{non-s}},$ where $p^I_{non-s}$ represents the final rumor density without supervisors. It is a metric that measures the degree of improvement in supervision compared to the unsupervised situation.
    \item Cost-effectiveness($C_e$): $C_e=\frac{C_E}{N_r},$ and let $C_e=1$ without supervisors to be a comparison. This metric measures the efficiency contribution of each supervisor.
    \item Marginal Benefit($M$): $M=\Delta C_E,$ that is, the current control efficiency of the $i$-th supervisor minus the control efficiency of the ($i-1$)-th supervisor. This metric is used to evaluate the change in gains contributed by each supervisor.
\end{enumerate}

In the six scenarios shown in Fig.~\ref{fig:comprehensive_eval}(a), the final rumor density consistently decreases as the number of supervisors increases. Moreover, it is worth noting that the smaller the network size and the lower the value of $\beta$, the more significant the decrease in density, indicating stronger supervision effectiveness. Besides, Fig.~\ref{fig:comprehensive_eval}(b) shows that control efficiency improves as the number of supervisors increases, approaching 1 in most cases. It can be observed that only 10 supervisors are needed to achieve approximately 0.8 efficiency for $N = 400$ and $\beta_1= 0.2$, while a much larger number is required to reach the same efficiency for $N = 1000$ and $\beta_1= 0.6$. This difference further reflects the combined influence of network size and rumor transmission rate on the speed of supervision efficiency improvement. In Fig.~\ref{fig:comprehensive_eval}(c), the result shows that cost-effectiveness decreases as the number of supervisors increases, especially in a larger network. We argue this indicates that, in reality, resource allocation should seek a reasonable threshold to maximize cost-effectiveness, rather than incurring wasted resources through reckless input. Turn to Figs.~\ref{fig:comprehensive_eval}(d) and~\ref{fig:comprehensive_eval}(e), we find that the marginal control benefit brought by each additional supervisor decreases as the number of supervisors increases in a smaller network. But when the network is larger, the marginal benefit tends to first increase and then decrease, with a certain degree of volatility. However, generally speaking, the smaller the network and the larger the $\beta$, the faster the marginal benefit decreases as the number of regulators increases. When $N = 400$ and $\beta_1=0.2$, the initial marginal benefit per supervisor is about 0.08, but it approaches 0 after more than 25 supervisors, suggesting that early deployment contributes more significantly than later additions. This also indicates that in practical applications, early resource investment should be prioritized over late-stage remediation. Then, as shown in Fig.~\ref{fig:comprehensive_eval}(f), we further quantify the sensitivity of network size and the impact of rumor spread. The results show that the smaller the network size, the lower the rumor propagation rate, and the higher the average control efficiency, which once again confirms the value of advocating the adjustment of supervision resource deployment strategies according to scenarios. Finally, combined with the results of Tab.~\ref{tab:network_scale_transmission_performance} and Fig.~\ref{fig:comprehensive_eval}, we determine the optimal supervisor configuration in different scenarios. For example, when $N = 400$, $\beta_1 = 0.2$, only 20 supervisors can achieve full control efficiency; When $N = 1000$, $\beta_1 = 0.6$, the 80 supervisors can only achieve partial efficiency.

In Figs. \ref{fig:caltech}(a)--(c), the three subfigures correspond to $\beta= 0.12, 0.25,0.6$, respectively. At a low transmission rate ($\beta= 0.12$, Fig.~\ref{fig:caltech}(a)), the density of I nodes rises to a moderate peak and then declines quickly to a low steady-state level, while Q nodes accumulate to become the dominant state in the network. This indicates that 50 supervisors are sufficient to effectively contain rumor spreading when the transmission intensity is weak. As $\beta$ increases to 0.25 (Fig.~\ref{fig:caltech}(b)), the peak of I-node density becomes higher and the decline is slower, with the steady-state I-node density settling at a noticeably higher level. The density of S nodes drops more rapidly, and both Q and R nodes occupy larger proportions of the network. When $\beta$ is further raised to 0.6 (Fig.~\ref{fig:caltech}(c)), the I-node density surges to a very high peak within the first few time steps and remains at an elevated steady-state level, while S nodes are almost entirely depleted. This demonstrates that under high transmission intensity, even with 50 supervisors, the supervision mechanism struggles to keep pace with the explosive rumor spreading in the real-world network structure, consistent with the findings from synthetic BA networks.

In Figs.~\ref{fig:caltech}(d)--(f), with $\beta$ fixed at 0.2, the three subfigures show the results for 0, 60, and 120 supervisors, respectively. Without any supervisors (Fig.~\ref{fig:caltech}(d)), the rumor spreads uncontrollably: I-node density rises sharply and stabilizes at a high level, while S nodes diminish rapidly and Q nodes are generated solely through natural state transitions (R $\rightarrow$ Q), remaining at a low density. When 60 supervisors are deployed (Fig.~\ref{fig:caltech}(e)), a significant suppression effect is observed---the peak of I-node density is substantially reduced, the decline after the peak is faster, and the steady-state I density is markedly lower. The Q-node density rises considerably, reflecting the supervisors' effectiveness in converting both S and I nodes toward less harmful states. With 120 supervisors (Fig.~\ref{fig:caltech}(f)), the suppression effect is further amplified: the I-node peak is even lower, the system reaches equilibrium more quickly, and the steady-state is dominated by Q and R nodes with only a minimal residual I-node density. These results on the Caltech Facebook network validate that the proposed model captures realistic rumor-supervision dynamics in real social network topologies, and that the key observations from synthetic networks---namely, the critical roles of transmission intensity and supervisor quantity---hold consistently in empirical settings.

\section{Conclusion and Outlook}
In this paper, we construct a novel confrontation framework to formalize the dynamic relationship between rumor propagation and supervisory intervention in complex networks. The core contribution of this work lies in advancing existing research through two main innovations. First, we extend the classical SIRS rumor propagation model by introducing a vigilant state Q and confrontation mechanisms, thereby enabling a more refined description of individual states and their behavioral responses during rumor spreading. Second, we design a supervisor mobility strategy based on random walks driven by node propagation importance, which allows supervisors to more effectively target rumor spreaders and high-risk individuals. Together, these mechanisms provide a finer-grained representation of the competitive dynamics involved in rumor containment.

Through simulations across different network topologies and parameter settings, we systematically examine how multiple factors, including network size, initial rumor density, propagation rate, supervisor allocation, and confrontation strength, jointly shape the overall system behavior. The results show that supervision effectiveness is governed by the interaction between structural and dynamical factors. In general, smaller networks and lower propagation rates lead to higher control efficiency, while earlier deployment of supervisors yields substantially greater marginal benefits. Moreover, based on multiple evaluation indicators, we identify optimal supervisor allocation schemes under different network sizes and propagation intensities, thereby providing practical guidance for resource allocation in rumor-control scenarios. Our results also demonstrate that network topology, especially the hub structure of scale-free networks, imposes important constraints on supervision coverage and strategic mobility in confrontation dynamics. Finally, we validate the applicability of the proposed mechanisms on a real-world network by simulating rumor propagation on the Facebook network of the California Institute of Technology. Overall, these findings underscore the importance of rumor supervision in confrontation settings and show that combining supervision objectives with node propagation importance can support scenario-specific and proactive resource allocation strategies.

This study also points naturally to several directions for future research. First, the current framework could be extended to incorporate adaptive rumor strategies, multilayer networks, and temporal network structures. Second, introducing heterogeneous supervisor capabilities would further improve the realism of the model. Third, empirical validation based on real propagation data would strengthen the practical relevance of the framework and help assess its predictive power in real-world applications. Finally, exploring the roles of individual behavioral heterogeneity and cognitive factors in confrontation mechanisms represents an important direction for deepening our understanding of rumor dynamics and their effective control.

\begin{acknowledgments}
This work was supported in part by the National Natural Science Foundation of China (NSFC) under Grant No.12272282 and in part by the Natural Science Foundation of Chongqing under Grant No. CSTB2025YITP-QCRCX0007. M.P. additionally acknowledges supported by the Slovenian Research and Innovation Agency under Grant No. P1-0403.
\end{acknowledgments}

\end{document}